\documentclass[fleqn,twoside]{article}
\usepackage{espcrc2}

\usepackage{epsf}
\usepackage[figuresright]{rotating}

\newcommand{\AmS}{{\protect\the\textfont2
  A\kern-.1667em\lower.5ex\hbox{M}\kern-.125emS}}

\title{New prospects for polarized hyperon fragmentation functions}

\author{J. Soffer\address{Centre de Physique Th\'eorique, CNRS Luminy Case 907,\\
F 13288 Marseille Cedex 09, France}}

\begin{document}

\begin{abstract}
We study the inclusive production of $\Lambda$ ($\bar \Lambda$) in several
high energy collision processes ($e^+e^-,~e^{\pm}p,~\nu (\bar \nu)p,~pp$), in view of an
accurate determination of the unpolarized and polarized fragmentation functions 
of a quark into a $\Lambda$ ($\bar \Lambda$). 
For polarized fragmentation functions the inaccuracy and the scarcity of present data
do not allow to distinguish between various theoretical models. We will indicate 
how future measurements will provide ways to discriminate between them and 
also how to achieve a necessary quark flavor separation. A possible extension
to the production of the other hyperons ($\Sigma^{\pm,0}, \Xi^{-,0}$), will be also briefly discussed.

\vspace{1pc}
\end{abstract}

\maketitle

\section{Introduction}
The usual parton $(f=q,\bar q,g)$ distribution functions $f_H(x,Q^2)$ of a hadron $H$, 
which are extracted from Deep Inelastic Scattering (DIS) for a space-like energy scale
$(Q^2 \leq 0)$, have counterparts in the time-like region $(Q^2 \geq 0)$, which are the 
hadron fragmentation functions $D_f^H(z,Q^2)$. They represent the probability to find 
the hadron $H$ with a fraction $z$ of the momentum of the parent parton
$f$, at a given value of $Q^2$. Like distribution functions, fragmentation 
functions are {\it universal}
, that is process-independent, and likewise, their $Q^2$ dependence is fully predicted
by the QCD evolution equations. They can be measured in various high
energy collision processes namely, $e^+e^-$ annihilation, $e^+e^- \rightarrow
(\gamma^{*}, Z) \rightarrow H X$, semi-inclusive DIS, $lp \rightarrow l' H X$ and
single-inclusive $pp$ collisions, $pp \rightarrow H X$. So far, the best known 
fragmentation funtions are $D^{\pi}_f$ and $D^K_f$, corresponding to the most
copiously produced light mesons $\pi$ and $K$. Our purpose here is to consider
the hyperon fragmentation functions and more specifically to examine the case
of $\Lambda$ ($\bar \Lambda$). The self-analyzing properties of  $\Lambda$
($\bar \Lambda$) make this spin-half baryon particularly appealing, to study
spin transfer mechanisms.
We first recall the results of a QCD analysis of the data for inclusive 
($\Lambda + \bar \Lambda$)
production in $e^+e^-$ collisions, which
yields the first simple and reliable parametrization of the unpolarized 
fragmentation functions
$D^{\Lambda, \bar \Lambda}_f(z,Q^2)$. The observed longitudinal polarization 
of the $\Lambda$'s produced
at LEP on the $Z$-resonance, leads to some inaccurate information on the
spin-dependent fragmentation
functions $\Delta D_f^{\Lambda}(z,Q^2)$. As we will see, several theoretical 
models have been proposed for these
polarized fragmentation functions which are, so far, badly constrained by 
existing data. Some predictions
can be made for the spin transfer in polarized DIS, 
but one gets no definite conclusion
by comparing them with the present very poor data from HERMES at DESY and
E665 at FNAL. We also stress the importance of the
$\Lambda$ ($\bar \Lambda$) production in neutrino (antineutrino) DIS, which allows a clean
flavor and spin separation. We will show some new data from NOMAD at CERN
and comment on future experimental possibilities.\\
We will also give the prospects from $pp$ collisions with polarized protons 
at RHIC-BNL, because there are
recent interesting suggestions for measuring the helicity (and transversity)
transfer asymmetry in the
process $p \overrightarrow p \rightarrow \overrightarrow \Lambda X$. 
From its dependence on the rapidity of the 
$\Lambda$, it is possible to discriminate easily between the various
theoretical models, thanks to the high luminosity
and the small statistical errors. Of course, this can be easely extended to the production
of the other hyperons ($\Sigma^{\pm,0}, \Xi^{-,0}$), which will be also briefly discussed.

\section{ $e^+e^-$ collisions}

From the production of $\Lambda$ ($\bar \Lambda$) in $e^+e^-$ annihilation one can
determine  $ D_f^{\Lambda}(z,Q^2)$ by using a standard parametrization,
at initial scale $\mu^2$
\begin{equation}
D_f^{\Lambda}(z,\mu^2)=N_f z^{\alpha_f}(1-z)^{\beta_f}~.
\end{equation}
To reduce the number of free parameters occuring above, one must make some simplifying 
assumptions based on, {\it e.g.} $SU_f(3)$ arguments.
{\it Ten} free parameters, needed to include light quarks, heavy quarks and gluons, 
were obtained by fitting the world data for
14GeV$\leq \sqrt s \leq $91.2GeV, in a leading and next-to-leading order (LO and
NLO) QCD analysis \cite{FSV}. Although, it leads to a good determination of $D_f^{\Lambda}$, 
there is no flavor separation for the dominant contributions $D_q^{\Lambda} (q=u,d,s)$ and no
separation between $D_q^{\Lambda}$ and $D_{\bar q}^{\Lambda}$, since the data
does not separate $\Lambda$ and ($\bar \Lambda$). Other sets of quark fragmentation
functions into $\Lambda$ ($\bar \Lambda$), based on a bag model calculation and a fit to $e^+e^-$
data, have been published recently \cite{BLT}.

The fragmentation of a longitudinally polarized
parton into a longitudinally polarized $\Lambda$ will be described in terms of
\begin{equation}
\Delta D_f^{\Lambda}(z,Q^2)= D_{f(+)}^{\Lambda(+)}(z,Q^2)- D_{f(+)}^{\Lambda(-)}(z,Q^2)~,
\end{equation}
where  $D_{f(+)}^{\Lambda(+)}(z,Q^2)$ [$D_{f(+)}^{\Lambda(-)}(z,Q^2)$] is the
probability to find a $\Lambda$ with positive [negative] helicity in a parton
$f$ with positive helicity. Clearly one obtains the unpolarized fragmentation
function $D_f^{\Lambda}$ by taking in (2), the sum instead of the difference.
For the fragmentation of a transversely polarized parton into a transversely
polarized $\Lambda$, one uses (2) to define $\Delta_T D_f^{\Lambda}(z,Q^2)$
in complete analogy with $\Delta D_f^{\Lambda}(z,Q^2)$.

\begin{figure}%[htb]
\begin{center}
\leavevmode {\epsfysize=6cm \epsffile{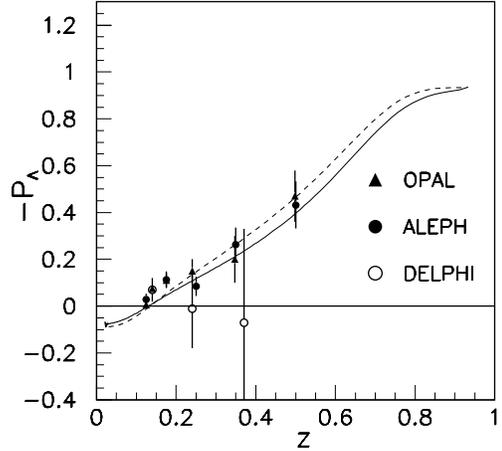}}
\end{center}
\caption[*]{\baselineskip 5pt Comparison of the LEP data
for ($-P_{\Lambda}$), at the $Z$-pole, with the theoretical
calculations of the SU(6) quark-diquark model (Taken from Ref.\cite{MSY}).}
\label{Fig1}
\end{figure} 

In order to construct $\Delta D_f^{\Lambda}$, one 
usually makes the assumption
\begin{equation}
\Delta D_f^{\Lambda}(z)=C_f^{\Lambda}(z)D_f^{\Lambda}(z)~,
\end{equation}
at a given scale, where $C_f^{\Lambda}(z)$ are the spin transfer coefficients, for 
which several models can be found in the literature. In Ref.\cite{FSV},
$\Delta D_f^{\Lambda}$ was neglected for heavy quarks and gluon and they take
the following simple parametrization
\begin{equation}
\Delta D_u^{\Lambda}=\Delta D_d^{\Lambda}=N_u \Delta D_s^{\Lambda}~~\mbox{and}~~
\Delta D_s^{\Lambda}=z^{\alpha}D_s~,
\end{equation}
in terms of {\it two} new parameters $\alpha$ and $N_u$ at a given scale. To cover a 
wide range of plausible models, they consider three different scenarios:\\
1) In the non-relativistic quark model, only strange quarks can contribute, therefore $N_u=0$.\\
2) By transposing the estimate of Ref.\cite{BJ} for the polarized $u$ and $d$
quark distributions in a $\Lambda$, to the polarized fragmentation functions, 
one gets $N_u=-0.2$.\\
3) By assuming that all polarized fragmentation functions are equal, which is
an extreme case, one gets $N_u=1$.\\
In Ref.\cite{BLT}, only two scenarios similar to 1) and 2), were considered and
for other alternatives along the same lines, see also Ref.\cite{KBH}. A different way
to construct $\Delta D_q^{\Lambda}(z)$ was proposed in Ref.\cite{MSY}, based on our
knowledge of the quark distributions and by means of the simple reciprocity relation
, the so called Gribov-Lipatov relation \cite{GLR}
\begin{equation}
D_q^H(z) \sim q_H(x),
\label{GLR}
\end{equation}
where $q_H(x)$ is the quark
distribution for finding the quark $q$
carrying a momentum fraction $x=z$ inside the hadron $H$.\\
The longitudinal $\Lambda$ polarization $P_{\Lambda}$, which can be directly
expressed in terms of the $\Delta D_f^{\Lambda}~'s$ (see ref.\cite{FSV}), has been 
measured at LEP by several experiments and is shown in Fig.1. It is large and negative
and consistent with the SU(6) quark-diquark model proposed in Ref.\cite{MSY}, as well as 
scenario 3) from Ref.\cite{FSV}, mentioned above. In a very recent work \cite{MSSY11}, 
one finds various predictions for $P_B$, where $B$ stands for any hyperon of the octet
baryon, obtained by using SU(3) symmetry relations.

\section{Charged lepton DIS process}

 These results can be used to make predictions for $\Lambda$ production in
semi-inclusive DIS, with only the electron (muon) beam longitudinally polarized. The
$\Lambda$ polarization $P_{\Lambda}$ along its own momentum is related 
(~for the exact expression see Ref.\cite{MSSY5})
to the longitudinal spin transfer $A^{\Lambda}(x,z)$ defined as
\begin{equation}
A^{\Lambda}(x,z)= {\sum_q e^2_q q^N(x)\Delta D^{\Lambda}_q(z) + (q \rightarrow \bar q)
 \over \sum_q e^2_q q^N(x) D^{\Lambda}_q(z) + (q \rightarrow \bar q)}~.
\end{equation}
From the above values of $N_u$, one anticipates that $A^{\Lambda}$ will be small
for 1), negative for 2) and large and positive for 3) and the results
can be found in Ref.\cite{FSV}. Another set of predictions is given 
in Ref.\cite{KBH} and it will be interesting 
to compare them with data from future polarized DIS experiments,
in particular from HERMES at DESY and COMPASS at CERN.

For the time being, we only have the few data points displayed in
Fig.2 for the spin transfer for $\Lambda$ production and in Fig.3
for $\bar \Lambda$ production. The three curves represent the predictions 
from three possible scenarios proposed in Ref.\cite{MSSY5} in the framework
of the pQCD based model and the use of the reciprocity relation Eq.(5). 
The dotted curves correspond to pure
valence quark contributions and for the dashed curves one has 
introduced a symmetric quark-antiquark sea, whereas it is asymmetric
for the solid curves. Although this last case seems slightly favored, the
simultaneous description of $\Lambda$ and $\bar \Lambda$ spin transfers
remains hard to achieve and therefore more accurate data are badly needed  
(for a detailed discussion on this point, see Ref.\cite{MSSY5} ).

\begin{figure}%[htb]
\begin{center}
\leavevmode {\epsfysize=5.8cm \epsffile{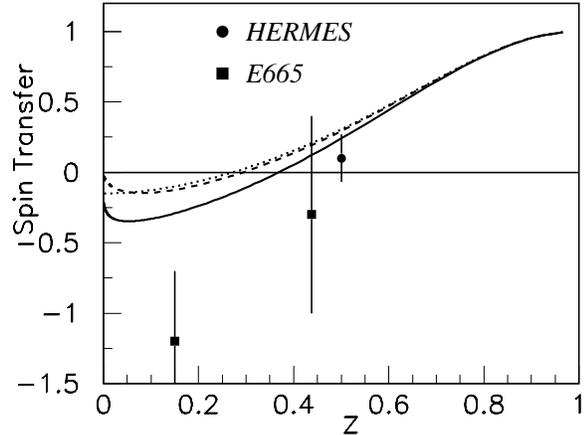}}
\end{center}
\caption[*]{\baselineskip 5pt Three different predictions for the
$z$-dependence of the $\Lambda$
spin transfer in charged lepton DIS, compared with data (Taken from Ref.\cite{MSSY5}).}
\label{Fig2}
\end{figure}

\begin{figure}%[htb]
\begin{center}
\leavevmode {\epsfysize=5.8cm \epsffile{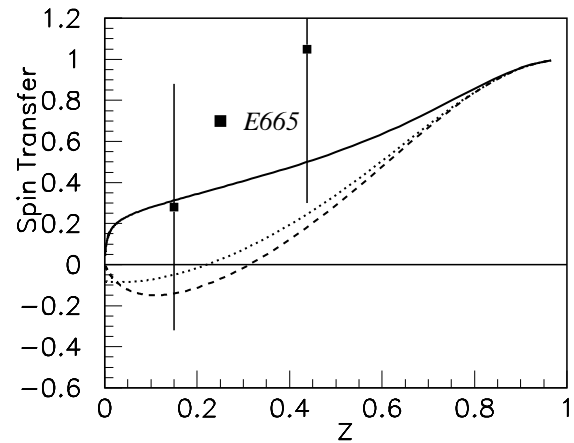}}
\end{center}
\caption[*]{\baselineskip 5pt Three different predictions for the
 $z$-dependence of the $\bar\Lambda$
spin transfer in charged lepton DIS, compared with data (Taken from Ref.\cite{MSSY5}).}
\label{Fig3}
\end{figure}

\section{Neutrino DIS process}
The main observation is based on the fact that neutrinos (antineutrinos) can be regarded
as a purely polarized lepton beam since neutrinos are left-handed (antineutrinos are
right-handed). Therefore they only interact with quarks of specific helicities and 
flavors. The scattering of a neutrino (antineutrino) beam on a hadronic target provides
{\it a source of polarized quarks with specific flavor}. This important property makes 
neutrino (antineutrino) DIS, an ideal laboratory to study the flavor dependence of
quark fragmentation.\\
If one measures, $\Lambda$ and $\bar \Lambda$ production in neutrino and antineutrino
 DIS, one can show \cite{MS} that the four longitudinal polarizations $P_{\nu}^{\Lambda}$, 
 $P_{\bar \nu}^{ \Lambda}$, $P_{\nu}^{\bar \Lambda}$ and $P_{\bar \nu}^{\bar \Lambda}$ can 
 be expressed in terms of four fragmentation functions, allowing a clean separation
 between quarks and antiquarks. In Fig.4 we show the NOMAD data for $P_{\nu}^{\Lambda}$, 
 compared with some model calculations from Ref.\cite{MSSY11}, as well as the predictions
 for the remaining three processes. Again more data are requiered to test these models
 but fortunately, one can expect future experimental possibilities from charged currents
 $e^{\pm}p$ DIS, with HERA at DESY, CEBAF at Jefferson Lab. and eRHIC at BNL.
 
\begin{figure}%[htb]
\begin{center}
\leavevmode {\epsfysize=5.5cm \epsffile{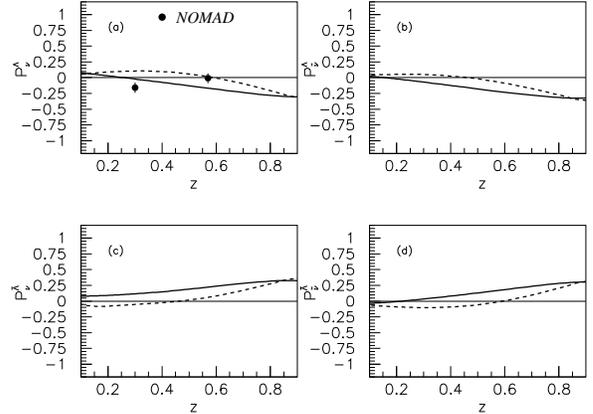}}
\end{center}
\caption[*]{\baselineskip 5pt Two different predictions for the
$z$-dependence of the $\Lambda$ and $\bar \Lambda$ polarization
in neutrino and antineutrino DIS (Taken from Ref.\cite{MSSY11}).}
\label{Fig4}
\end{figure}

\begin{figure}%[htb]
\begin{center}
\leavevmode {\epsfysize=7.5cm \epsffile{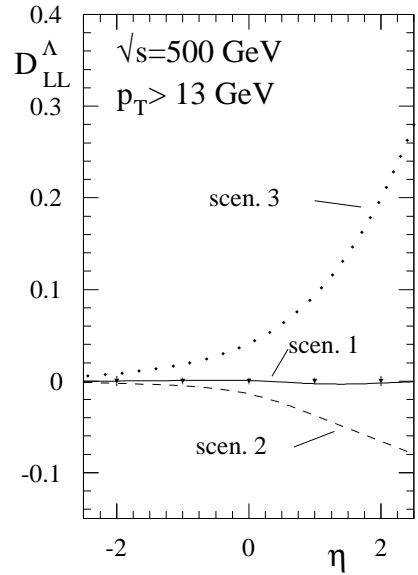}}
\end{center}
\caption[*]{\baselineskip 5pt The helicity transfer parameter $D_{LL}^{\Lambda}$,
versus the rapidity of the $\Lambda$ at RHIC-BNL maximum energy, using  
three different scenarios described in Section 2 (Taken from Ref.\cite{FSVa}).}
\label{Fig5}
\end{figure}

\begin{figure}%[htb]
\begin{center}
\leavevmode {\epsfysize=8.5cm \epsffile{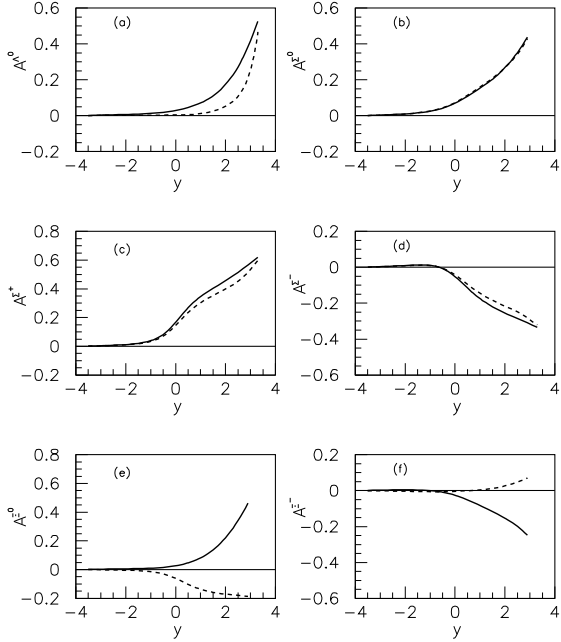}}
\end{center}
\caption[*]{\baselineskip 13pt The helicity transfers, denoted here $A^B$, as functions of
rapidity $y$ of the produced $\Lambda,\Sigma^{\pm,0}, \Xi^{-,0}$, octet
baryon members, in $\vec{p}p$ collisions at
$\sqrt{s}=500~\rm{GeV}$, with the spin-dependent fragmentation
functions in the pQCD counting rules analysis (solid curves) and
the SU(6) quark-diquark spectator model (dashed curves). Note that
the dashed and solid curves in (b) almost overlap ( Taken from Ref.\cite{MSSY9}.}
\label{Fig6}
\end{figure}
\section{Hyperons in $pp$ collisions}
Let us now move to another area where we can test these polarized fragmentation functions, 
by studying the spin transfer parameters in $pp$ collisions. A new polarized
$pp$ collider at RHIC-BNL, which starts operating now, will allow to undertake 
a vast spin physics programme at high
center-of-mass energies, up to $\sqrt s$ = 500GeV. First, we are interested in the
reaction $\overrightarrow{p}p  \rightarrow \overrightarrow{\Lambda} X$, where both the
initial proton and the produced $\Lambda$ are longitudinaly polarized and we consider the
helicity transfer parameter $D^{\Lambda}_{LL}$ defined as follows
\begin{equation}
D^{\Lambda}_{LL}(\eta,p_T)=(d\sigma_{++}-d\sigma_{+-}) /(d\sigma_{++}+d\sigma_{+-})~,
\end{equation}
\begin{figure}[ht]
\begin{center}
\leavevmode {\epsfysize=8cm \epsffile{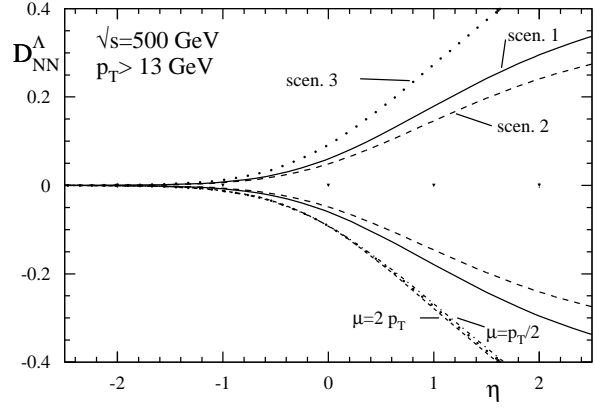}}
\end{center}
\caption[*]{\baselineskip 5pt  Upper and lower bounds for the spin transfer
parameter $D_{NN}^{\Lambda}$, versus the rapidity of the $\Lambda$ at
RHIC-BNL maximum energy. The small statistical errors are shown
along the $\eta=0$ axis (Taken from Ref.\cite{FSSV}) .}
\label{Fig7}
\end{figure}

where $\eta$ and $p_T$ are the rapidity and the transverse momentum of the outgoing $\Lambda$. 
$d\sigma_{++}$ [$d\sigma_{+-}$] denotes the cross section
where the proton and the $\Lambda$ helicities have the same [opposite] sign. When $\eta$
is positive the $\Lambda$ has the direction of $\overrightarrow{p}$ and $p_T$
is always assumed to be large. $D^{\Lambda}_{LL}$ is directly related to the $\Delta D_f^{\Lambda}~'s$,
whose exact expression can be found in Ref.\cite{FSVa}, and therefore 
it is possible to make predictions, which are shown in Fig.5. 
Again one can see that the sign and magnitude of  $D^{\Lambda}_{LL}$  are strongly correlated to the
values of $N_u$ and given the smallness of the expected statistical errors, indicated in
the figure along the horizontal axis, it will be now rather easy to pin down the right 
underlying mechanism.\\
One notices that in all cases $D^{\Lambda}_{LL}$ is small for negative rapidity. This is 
due to the fact that the dominant subprocess is $gq \rightarrow gq$ and we have assumed that
$\Delta D_g^{\Lambda} = 0$. This feature is rather general and it was discussed in more details
in Ref.\cite{MSSY9}. It characterizes also other helicity transfers as shown in Fig.6,
where we display the predicted helicity transfers for all the hyperons $\Lambda,\Sigma^{\pm,0}, \Xi^{-,0}$, 
resulting from two models of the fragmentation functions.

Finally, we can also consider the case where both the proton and the 
$\Lambda$ are transversely polarized and
the corresponding spin transfer parameter $D_{NN}^{\Lambda}$, is
defined as in Eq.(7), where helicity is replaced by transverse polarization.
This spin observable is related to the quark transversity distributions,
so called $h_1^q(x,Q^2)$ and to the transversity fragmentation functions
$\Delta_T D_f^{\Lambda}(z,Q^2)$ mentioned in Section 2. Needless to say that we
have no experimental information on these quantities. However we
can derive some bounds on $D_{NN}^{\Lambda}$, using positivity arguments.
We recall that positivity leads to the inequality \cite{js}
\begin{equation}
2|h_1^q(x,Q^2)| \leq q(x,Q^2) + \Delta q(x,Q^2)~,
\end{equation}
whose validity has been established up to NLO QCD corrections.
A similar constraint holds for hadron fragmentation functions \cite{FSSV},namely
\begin{equation}
2|\Delta_T D_q^H(z,Q^2)| \leq D_q^H(z,Q^2) + \Delta D_q^H(z,Q^2)~.
\end{equation}
By saturating the above inequalities, we get an estimate for an upper and lower bounds 
for $D_{NN}^{\Lambda}$ which are displayed in Fig.7. The larger allowed
region corresponds to scenario 3), mentioned earlier, due to the value of $N_u$ and from the expected
statistical accuracy, we get rather small error bars.\\

To summarize, we have now some knowledge of the unpolarized fragmentation 
functions $D^{\Lambda}_f(z,Q^2)$, but
 we miss a better flavor separation, which can be properly achieved 
by means of neutrino (antineutrino)
DIS or charged currents in $e^{\pm}p$ DIS. 
The corresponding polarized fragmentation functions
$\Delta D^{\Lambda}_f(z,Q^2)$ are poorly known and more data is needed for testing
several theoretical models proposed in the literature. Future measurements of $\Lambda$ and 
$\bar \Lambda$ production in semi-inclusive charged lepton DIS will improve the situation
and this can be generalized to the case of the production of the 
other hyperons $\Sigma^{\pm,0}, \Xi^{-,0}$. The new polarized $pp$ collider at RHIC-BNL, with 
high luminosity and high energy, is also a very powerful facility to study unpolarized and
polarized fragmentation functions for all octet baryon members.

{\it Acknowledgements}

It is my pleasure to thank the orginazers S.D.Bass, A. De Roeck and A. Deshpande for their
invitation and for setting up this excellent workshop in such pleasant and stimulating
atmosphere.

\end{document}